\documentclass[pre,twocolumn]{revtex4-1}
\usepackage{graphicx,subfigure,hyperref}
\usepackage{epstopdf}
\expandafter\let\csname equation*\endcsname\relax
 \expandafter\let\csname endequation*\endcsname\relax
\usepackage{amssymb,amsmath}
\graphicspath{%
    {converted_graphics/}
    {/}
}
\usepackage{soul, color}

\begin{document}
\title{A Cooperative Sequential Adsorption Model in Two Dimensions with Experimental Applications for Ionic Self-Assembly of Nanoparticles}
\author{L. Jonathan Cook, D. A. Mazilu, I. Mazilu, B. M. Simpson, E. M. Schwen, V. O. Kim, A. M. Seredinski}
\affiliation{ Department of Physics and Engineering, Washington and Lee University, Lexington, VA 24450}

\begin{abstract}
Self-assembly of nanoparticles is an important tool in nanotechnology, with numerous applications including thin films, electronics, and drug delivery.  We study the deposition of ionic nanoparticles on a glass substrate both experimentally and theoretically.   Our theoretical model consists of a stochastic cooperative adsorption and evaporation process on a two-dimensional lattice.  By exploring the relationship between the initial concentration of nanoparticles in the colloidal solution and the density of particles deposited on the substrate, we relate the deposition rate of our theoretical model to the concentration.
\end{abstract}

\maketitle



\section{Introduction}\label{intro}

Self-assembly of nanoparticles is an important tool in nanotechnology, and an active area of interdisciplinary research, with applications spanning a variety of fields such as optics, materials science, electronics, and nanomedicine \cite{book on nano}. There are a multitude of  experimental techniques that take advantage of the natural tendency of particles to self-assembly due to chemical bonds \cite{book on nano}. One of the most intuitive and cost-effective methods for creating thin films, for example, is layer-by-layer self-assembly, also known as ionic self-assembly of monolayers, or ISAM. It was first introduced by Iler \cite{iler} and later used in the creation of antireflective coatings \cite{lvov}. To create the thin films, layers of cations and anions are deposited by alternately dipping the substrate in aqueous solutions of appropriate ions. The same principle is used for other types of applications, such as the creation of thin films using nanotubes \cite{book on nano}, and drug attachment and encapsulation of nanoparticles using synthetic dendrimers \cite{dendri1, dendri2}.

The main goal of our paper is to present a stochastic statistical physics model that encapsulates the essential features of the self-assembly of ionic nanoparticles from a colloidal suspension onto a glass substrate. The model is validated by experiments conducted in our lab  that show  the effect of the concentration of the colloidal suspension on the nanoparticle coverage of a glass substrate.

Analytical and computational  models of sequential adsorption have proven successful in describing diverse physical systems ranging from surface deposition and chemisorption on crystal surfaces \cite{liggett} to epidemic problems \cite{diekmann,murray} and voting behavior \cite{voting}.  The dynamics of nanoparticle deposition is another application of such models.  It is currently an  active area of research in nanotechnology studies \cite{heflin} which addresses interesting open questions on the theoretical front \cite{gouet}.

Two classes of models that have been particularly successful are \it random sequential adsorption \rm  (RSA) \cite{evans, cadilhe, cadilhe1}, in which particles are adsorbed/deposited at a fixed rate at random unoccupied sites on a grid, and \it cooperative sequential adsorption \rm (CSA) \cite{privman}, in which adsorption rates depend upon the occupation of neighboring sites. One-dimensional sequential adsorption models have been studied thoroughly in different physical contexts \cite{evans, privman}, but adsorption in two dimensions is  less understood. There are many computational adsorption models \cite{redner}, but few analytical solutions have been developed for the general two-dimensional case. Recently, analytical results have been reported for the random sequential process \cite{cadilhe} and reaction-diffusion processes on Cayley trees and Bethe lattices \cite{ben avraham, matin, ali, mazilu1,tome,fonseca}. Adding  the possibility of particle detachment, or \textit{evaporation},  to such models brings additional complications.  One of the standard tools used to study these systems, the \emph{empty-interval method} \cite{redner}, fails when evaporation is considered. Evaporation has been treated analytically in a few studies of one-dimensional systems using a quantum mechanical approach \cite{grynberg}.

We will describe the ISAM process using a \textit{stochastic cooperative sequential adsorption with evaporation} (CSAE) model on two-dimensional lattices.  These methods are ideally suited for modeling ISAM since the deposition process of nanoparticles is stochastic and the deposited nanoparticles are electrically charged, as are the substrate deposition sites, suggesting a cooperative sequential adsorption model with deposition rates dependent on nearest-neighbor site occupation. 

The general CSAE model that we utilize leads to analytical solutions for the particle density. In order to validate the model, we conducted an experimental study of the effect of the colloidal suspension concentration on the steady state coverage density of the glass substrate. The analytical results are compared to the experimental data and to Monte Carlo computer simulations. We found excellent agreement between the three methods. 

In the following section of this paper, we describe the ISAM process and our experimental results for the concentration dependence of the coverage density. In section \ref{theory}, we present the analytical model and its mean field solution. Section \ref{comparison} is dedicated to a comparison between experiment, theory, and the Monte Carlo simulations. We summarize our results and discuss some open questions in section \ref{conclusion}.

\section{Experiment: ionic self-assembly of silica nanoparticles}\label{experiment}

\begin{figure}[htb]
\begin{center}
\includegraphics[width=8.6cm]{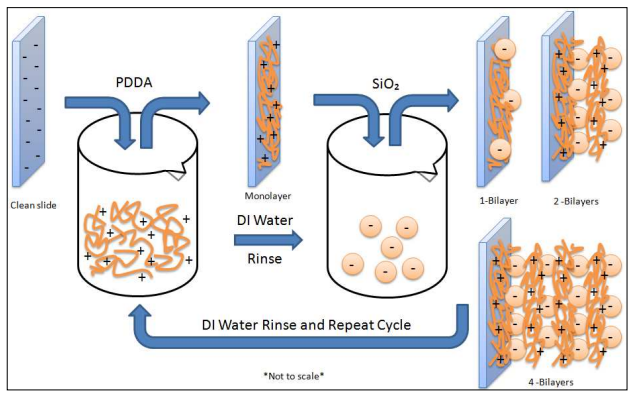}
\caption{(Color online) ISAM process of creating multiple bilayers.\label{fig1}}
\end{center}
\end{figure}

The use of ionic self-assembled monolayers allows detailed structural control of materials at the nanoscale, combined with ease of manufacturing and low cost.  The ISAM process allows the deposition of alternating layers of cations and anions by dipping a substrate in aqueous solutions of the appropriate ions, as illustrated in Fig.\ \ref{fig1}.  
Because it is a dipping process, any exposed surface is homogeneously coated, allowing highly uniform, conformal coatings on irregular shapes.  The basic building block for the film is a cation/anion bilayer, which may consist of either two polyelectrolytes (a polycation and a polyanion), a polyelectrolyte and a nanoparticle, or two different nanoparticles.  The thickness of a bilayer is a function of the diameter of the nanoparticle and the packing of the particles from layer to layer.  The optical properties of the resulting film can be tuned by the choice of nanoparticles and by the number of bilayers deposited.  A comprehensive review of the technique and its applications can be found in \cite{heflin}.  Although there are numerous studies on the subject of thin-film characterization \cite{films1,films2}, the goal of creating thin films with a graded index of refraction is still outstanding.  A study published by Yancey et al.\ \cite{ritter} shows that the coverage of the substrate plays an important role in tuning the index of refraction of the thin film.  The Maxwell-Garnett approximation \cite{optics book}, in fact, predicts that the index of refraction depends on surface coverage.

\begin{figure}[tbh]
\begin{center}
\includegraphics[width=8.6cm]{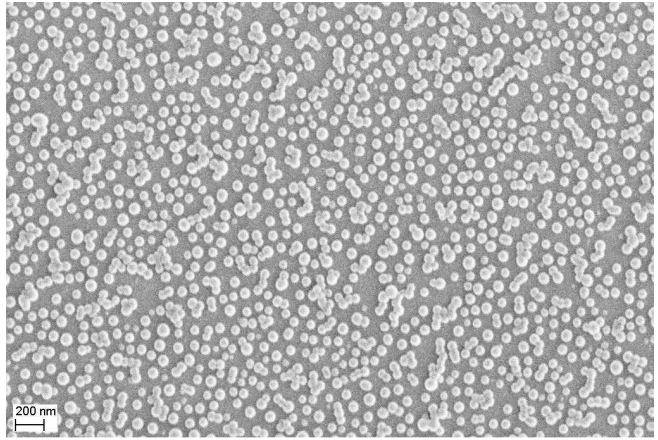}
\caption{A sample SEM micrograph at 25,000x magnification.  Nanoparticles deposited on the substrate appear as light regions.\label{fig2}}
\end{center}
\end{figure}

In our experiments we deposited negatively charged spherical silica nanoparticles of nominal 40-50 nm diameter on negatively charged glass slides using poly(diallyldimethylammonium chloride) (PDDA) as polycation. The silica nanoparticles (SNOWTEX ST-20L from Nissan Chemical) were in a colloidal suspension at stable $pH=10.3$ and room temperature $T=21^{\circ}C$. The glass slides were cleaned under sonication, in three successive twenty-minute steps, with LABTONE detergent, 1N sodium hydroxide solution, and deionized water, and then dried with flowing nitrogen gas. The dipping time was ten minutes for each bilayer. We varied the concentration of the silica suspension by diluting it with deionized water. We examined the nanoparticle coverage of the substrate using SEM micrographs, in which deposited particles appear as light regions on a dark background. A sample SEM micrograph is shown in Fig.\ \ref{fig2}.  
We processed two single-bilayer micrographs for each concentration data point. Using an automated pixel-counting method we determined the average coverage of light pixels, representing presence of deposited particles. The experimental data is presented in section \ref{comparison}, alongside and compared to the analytical solution.

\section{Theoretical model for the ionic self-assembly process}\label{theory}

We model the glass slide uniformly covered by the PDDA polymer as a  finite two-dimensional lattice with $N$ total sites. We consider the silica nanoparticles as charged monomers that attach to and detach from the lattice sites. We define an occupation number $n_{i}=0$ for an empty site and $n_{i}=1$ for an occupied site.  Our model considers both evaporation of monomers  and deposition of monomers  with rates dependent on the number of occupied neighbors.   

We define the following transition rate for the particle occupation:
\begin{equation}
c(n_{i}\rightarrow(1-n_{i}))=\gamma n_{i}+(1-n_{i})\alpha\beta^{\sum_{j \in NN}n_{j}}
\end{equation}
The first term in the transition rate is the evaporation term: if a particle is present, it will evaporate with probability $\gamma$. The second term describes the deposition of monomers. If the lattice cell is empty, a monomer will attach with a rate equal to $\alpha \beta^{\eta}$, where $\eta=\sum_{j \in NN}n_{j}$ is the number of occupied nearest neighbors. When $\eta=0$ (no nearest neighbors present), the intrinsic deposition rate is $\alpha$. The rate $\beta$ attenuates this intrinsic deposition rate based on the number of occupied nearest neighbors.

Given this transition rate, the number of particles on the lattice changes according to the following equation:
\begin{align}
\frac{\partial n_i}{\partial t}=-&\gamma n_i+(1-n_i)\alpha\beta^\eta
\end{align}

\subsection{Mean field solution}
 
In order to find the equation that governs the time dependence of the overall particle density, we take the ensemble average of $n_i$:
\begin{align}
\frac{\partial \langle n_i\rangle}{\partial t}=&-\gamma \langle n_i\rangle+\langle(1-n_i)\alpha\beta^{\eta}\rangle
\end{align}
Since this equation contains higher order correlations, we employ the mean field technique, which allows for the correlations to be approximated as:
\begin{equation}
\langle n_in_j\rangle=\langle n_i\rangle\langle n_j\rangle
\end{equation}
Using this approximation, we arrive at the following equation:
\begin{align}
\frac{\partial \langle n_i\rangle}{\partial t}=&-\gamma \langle n_i\rangle+(1-\langle n_i\rangle)\alpha\beta^{\langle\eta\rangle}
\end{align}
where $\langle \eta\rangle=\sum_{j\in NN}\langle n_j\rangle$.
 
In Fig.\ \ref{fig2}, we see that the size of a particle is much less than the size of the slide. As such, edge effects due to the finite size of the slide should be negligible in the interior of the slide. Additionally, we see a uniform distribution of particles on the slide. From these two observations, we find that the coverage density is independent of the location on the slide. Therefore, the average site density $\rho_i=\langle n_i\rangle$ is the same as the overall average coverage density $\rho=\sum_i\frac{\rho_i}{N}$:
\begin{align}
\rho_i=\rho
\end{align}
Using this information leads to a rate equation for the particle density:
\begin{equation}\label{density-time}
\frac{\partial \rho}{\partial t}=-\gamma \rho+(1-\rho)\alpha\beta^{4\rho}
\end{equation}
 
\begin{figure}[tbh]
\begin{center}
\includegraphics[width=8.6cm]{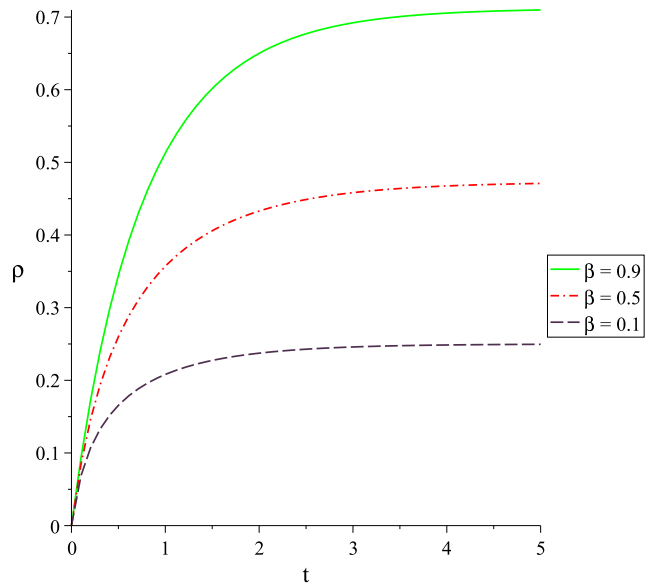}
\caption{(Color online) Particle density as a function of time for three values of $\beta$ with $\gamma=0.3$ and $\alpha=1$. The three values of $\beta$ shown are (black, dashed line) 0.1, (red, dot-dashed line) 0.5, and (green, solid line) 0.9.\label{fig3}}
\end{center}
\end{figure}

For the steady state, $\frac{\partial \rho}{\partial t}=0$, this is a self-consistent transcendental equation that can be solved numerically:
\begin{equation}\label{density-ss}
\rho=\frac{\alpha\beta^{4\rho}}{\gamma+\alpha\beta^{4\rho}}
\end{equation}
Eq.\ \eqref{density-time} can be solved numerically using standard software such as Maple or Mathematica. In Fig. \ref{fig3}, we present the time dependent particle density for three different values of $\beta$ (0.1, 0.5, and 0.9) at a fixed $\gamma=0.3$ and $\alpha=1$.
For all three $\beta$'s the shape of these curves is the same, but the steady state values of the particle density are drastically different. A more in-depth analysis of the steady state coverage is required and will be presented in section \ref{comparison}, in connection with its dependence on concentration.

From a theoretical point of view, it is worth noting  that Eq.\ \eqref{density-time} can be generalized for any coordination number $z$, as:

\begin{equation}\label{density-time-general}
\frac{\partial \rho}{\partial t}=-\gamma \rho+(1-\rho)\alpha\beta^{z\rho}
\end{equation}

The solution of this mean field equation can be obtained for a generic geometry with an effective coordination (per site)  $z$. Such a solution would show that a different coordination number (in the mean field limit) just corresponds to a different effective $\beta$, i.e., the parameter would be $\beta^z$ rather than $\beta$.  Because we are comparing our analytical results to a very specific experiment, we discus in this article just the  particular case of a square lattice of coordination number $z=4$ and $\beta<1$, to simulate electrostatic screening.

The case of $\beta>1$ corresponds to a physical situation in which the presence of occupied neighbors favors adsorption. This choice can apply to a voter- or an epidemic-type  model. The numerical solutions for  $\beta>1$  are very similar in shape to the ones  presented in Fig. \ref{fig3} for $\beta<1$. The main difference is the increased rate at which the lattice fills up to $100\%$ coverage. 
The cooperative aspects of the model disappear for $\beta=1$; it becomes a Langmuir model \cite{langmuir}.

\subsection{Connection to the Ising model}

We  arrive at the same self-consistent transcendental equation for the particle density  in the steady state if we approach the problem in the context of the Ising model on a two-dimensional lattice. The Hamiltonian associated with the  two-dimensional Ising model of a system of $N$ spins in an external field is:
\begin{equation}
H=-J\sum_{i,j\in NN}s_{i}s_{j}-B\sum_{i=1}^{N}s_{i}
\end{equation}

The first term describes the interactions between nearest-neighbor spins, and the second term expresses the interaction between each spin and an external magnetic field $B$. The spin numbers are $s_i=1$ for a spin up, and $s_{i}=-1$  for for a spin down.

We can map our model onto an equivalent spin model, in which the spin numbers are related to the occupation numbers $n_i$, $s_i=2n_i-1$. Using the detailed balance condition, we can solve for the coupling constants in terms of the defined attachment and detachment rates $\alpha$, $\beta$, and $\gamma$ :
\begin{eqnarray}
K=\frac{J}{kT}=\frac{1}{4}\ln(\beta)\\
h=\frac{B}{kT}=\frac{1}{2}\ln\left(\frac{\alpha\beta^2}{\gamma}\right)
\end{eqnarray}

 The equation derived for the magnetization of a system of spins in an external magnetic field in the mean field approximation is \cite{redner}:
\begin{equation}
M=\tanh(4KM+h)
\end{equation}
With the coupling constants $K$ and $h$ found above, and  $M=(2\rho-1)$, we arrive at a transcendental equation identical to Eq.\ \eqref{density-ss}.

\subsection{Computer simulations}
 
\begin{figure}[tbh]
\begin{center}
\includegraphics[width=8.6cm]{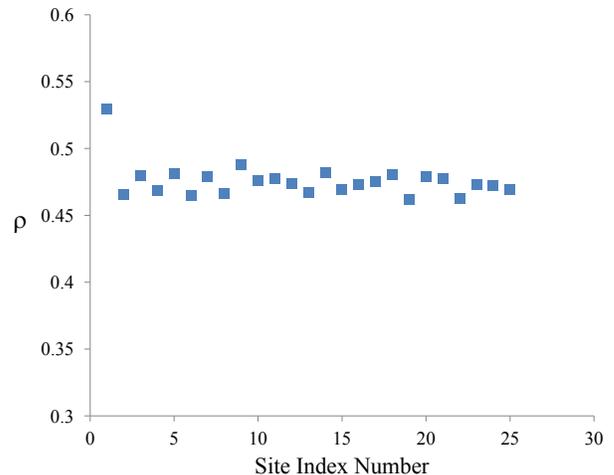}
\caption{Simulation results showing the the effects on the average site density near the left edge of the lattice. The sites are indexed with site 1 next to the edge.  The density shown is the column average of rows 31 to 90 on a $120\times120$ lattice with $\gamma=0.3$, $\alpha=1$, and $\beta=0.5$.\label{fig-edge}}
\end{center}
\end{figure}

We perform Monte Carlo simulations on a two-dimensional square lattice in order to investigate the dynamics of the CSAE process and evaluate the steady state solutions under various parameter regimes. Our simulations utilize a $120\times120$ two-dimensional grid onto which particles are both deposited at empty sites and evaporated from filled sites. In order to minimize edge effects, data is recorded for only the $100\times100$ matrix at the center of the larger lattice. The interior of the lattice is chosen instead of periodic boundary conditions; this choice mimics what is done experimentally. Only a small portion of the glass slide is analyzed, which is typically far away from the edges of the slide.  The edges in the simulations have a higher average site density due to the reduced number of neighbors.  Additionally, the average site density near the edges in the simulation decays rapidly to the average bulk density as shown in Fig.\ \ref{fig-edge} for the left edge.  Similar results are seen for the other edges.  Edge effects were not seen in the SEM micrographs, which is another reason we only consider the interior of the lattice.

Particles are deposited at empty cells with the rate $\alpha\beta^{\eta}$, where $\eta$ represents the sum of occupied neighboring sites. Particles evaporate from filled cells with the rate $\gamma$, which is independent of the state of neighboring sites. To update a site, a random site is chosen.  If the site is occupied, the particle will evaporate with rate $\gamma$.  If the site is empty, then it will become occupied with rate $\alpha\beta^\eta$.  To make the simulation reach the steady state more efficiently, we only consider updates that change the configuration of the lattice.  A different reaction (evaporation, adsorption with no neighbors, adsorption with one neighbor, etc.) is randomly chosen.  We weight the choice of each reaction by the number of sites which would allow this reaction multiplied by the rate of the reaction.  Once a reaction is chosen, a site with the given reaction is randomly chosen to change.  Starting with an empty lattice, we allow the system to reach steady state by waiting $1.44\times10^6$ site updates.  We then average the particle density at steady state over 100 realizations of the system.
 
For physical systems with repulsion between particles, the rate of deposition decreases when neighboring sites are occupied. We model this situation by choosing $\beta$ to be between zero and one for all simulations. A simple rescaling of time in Eq.\ \eqref{density-time} shows that the ratio of $\alpha$ to $\gamma$ controls the steady state density.  Therefore, we set $\alpha=1$ and vary $\gamma$ without loss of generality.
 
\begin{figure}[tbh]
\begin{center}
\subfigure[]{\includegraphics[width=8.6cm]{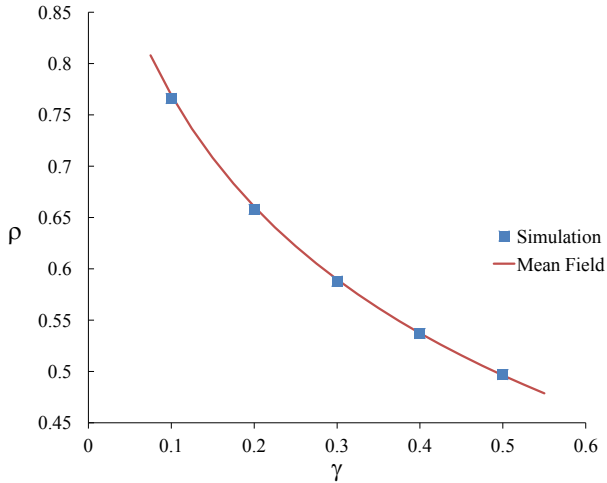}\label{fig5a}}
\subfigure[]{\includegraphics[width=8.6cm]{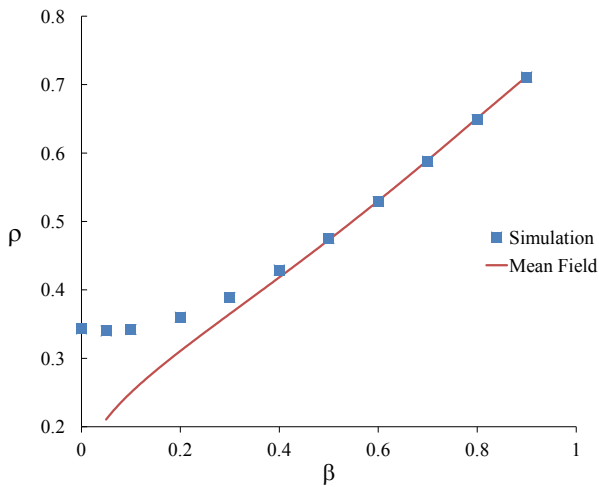}\label{fig5b}}
\caption{(Color online) Comparison of simulation (blue squares) and mean field results (red line). \subref{fig5a} Particle density as a function of $\gamma$ with $\alpha=1$ and $\beta=0.7$. \subref{fig5b} Particle density as a function of $\beta$ with $\alpha=1$ and $\gamma=0.3$.\label{fig5}}
\end{center}
\end{figure}

As seen in Fig.\ \ref{fig5}, the mean field result is in excellent agreement with the simulation results
and captures the relevant dynamics of this model for $\beta\ge0.4$. Additionally, we see in Fig.\ \ref{fig5b} that the density as a function of $\beta$ could be approximated by a linear function for $\beta\ge0.4$. At smaller values of $\beta$, the mean field theory fails to agree with the simulation due to stronger spatial correlation, which the theory neglects.  Since the simulation models an ISAM, we now look at connecting this linear dependence to a similar one found in the experimental data.

\section{Comparisons:  experiment and theory}\label{comparison}

Our CSAE model considers a simple case of deposition and evaporation of monomers and yields a transcendental equation for the particle density of the steady state that can be solved numerically. For the proposed model,  the equation \eqref{density-ss} associated with the steady state is re-written here for convenience:
\begin{equation}\nonumber
\rho=\frac{\alpha\beta^{4\rho}}{\gamma+\alpha\beta^{4\rho}}
\end{equation}

\begin{figure}[tbh]
\begin{center}
\includegraphics[width=8.6cm]{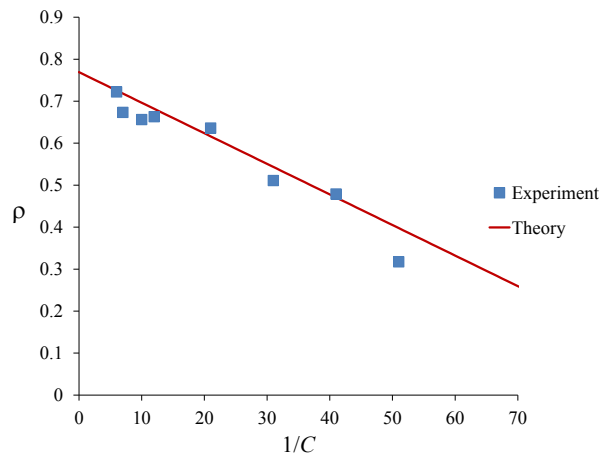}
\caption{(Color online) Comparison of experimental data and theory for particle density as a function of the inverse of concentration of the colloidal suspension in arbitrary units. The equations associated with the linear fit are: i) theory (red line), $y=-0.0077 x+0.7655$; ii) experiment (blue squares), $y=-0.0078 x+0.7566$ with $R^2=0.94681$.  The theoretical fit is drawn from the numerical solutions to Eq.\ \eqref{density-ss}.\label{fig6}}
\end{center}
\end{figure}

The experimental data shows a linear dependence between the particle density and the inverse of concentration. Using our model, we found a relationship between the concentration of the nanoparticle suspension and the theoretical probability rate $\beta$. In particular, we found numerical solutions for the particle density $\rho$ in Eq.\ \eqref{density-ss} for fixed $\alpha=1$ and $\gamma=0.3$, which match the experimental data shown in Fig.\ \ref{fig6}. Although equation \eqref{density-ss} is a nonlinear function, a linear approximation matches the numerical solution well, which is also shown in Fig.\ \ref{fig6}. 
From this comparison, we conclude that, according to our model, for constant temperature,  the concentration of the nanoparticle solution  is a function of $\beta$:
\begin{equation}\label{concentration}
C=\frac{1}{75(1-\beta)}
\end{equation}
for the chosen values of $\alpha=1$, and $\gamma=0.3$, or for any values of $\alpha$ and $\gamma$ for which the ratio $\frac{\gamma}{\alpha}=0.3.$

We linearize equation \eqref{density-ss} by performing a Taylor expansion about $\beta=1$.  
\begin{equation}
\rho=\rho(\beta=1)+(\beta-1)\left.\frac{\partial\rho}{\partial\beta}\right|_{\beta=1}+\dots
\end{equation}
where we only keep the linear term.
Using $\rho(\beta=1)=\frac{\alpha}{\gamma+\alpha}$, we obtain the following result:
\begin{equation}\label{theory-linear}
\rho=\frac{\alpha}{\gamma+\alpha}-(1-\beta)\left[4\left(\frac{\alpha}{\gamma+\alpha}\right)^2\left(1-\frac{\alpha}{\alpha+\gamma}\right)\right]
\end{equation}
As long as this ratio is equal to $0.3$, the intercept of the theoretical and experimental lines is identical, and the difference between the two slopes is minimal.

For the model presented in this paper, we also explored the case of the detachment rate being dependent on the number of neighbors, $\gamma^{\sum_{j \in NN}n_{j}}$. In the end, it seemed an unnecessary complication to consider such dependence, because the results showed that the model can be recast in terms of the ratio $\frac{\beta}{\gamma}$.
 
From an experimental point of view, this cooperative sequential adsorption model with evaporation can lead to interesting applications. The ability to predict or estimate the steady state coverage makes possible the prediction of the index of refraction \cite {optics book}, which is dependent on the overall particle density. A graded index of reflection is an outstanding goal in the creation of antireflective coatings. The model can also be modified for other lattice structures, such as Cayley trees with any coordination number $z$, with applications in modeling drug encapsulation in nanomedicine \cite{dendri1}.  The attachment and detachment rates can be chosen to reflect nearest-neighbor attraction ($\beta>1$) and repulsion ($\beta<1$).

\section{Conclusions}\label{conclusion}
In this article, we presented a cooperative sequential adsorption with evaporation model for the experimental process of ionic self-assembly of  charged silica nanoparticles (ISAM). Experimentally, we investigated the effect of the concentration of the nanoparticle suspension on the particle density at fixed temperature. We found that the particle density depends linearly on the inverse of the concentration. Theoretically, we found equations for the time-dependent particle density and the density of the steady state. Our solutions are also validated by Monte Carlo simulations. We compared our theoretical results with the experimental data, and found an excellent agreement between the two for $\beta\ge0.4$. We conclude that we can directly relate our theoretical probability rate $\beta$ to the nanoparticle concentration using Eq.\ \eqref{concentration}.

Several open questions may be addressed by extending our results.  The model presented matches well the particle density of the steady state found experimentally, but it doesn't capture the dynamics of the system on its way to the steady state. Experimental studies \cite{lvov} indicate that $90\%$ of the particle attachment happens in the first 30 seconds of the dipping process, followed by a slower approach to the final steady state. We plan to further study this time-dependent behavior both experimentally and theoretically. We will explore the dynamics of  our model when time-dependent attachment and detachment rates are being considered, in agreement with the experiment.  Our theoretical model can be generalized to include other aspects of the ISAM process, such as the presence of dimers and other particles of various shapes and sizes in the  colloidal suspension. The analytical method used in our study is the mean field method, which disregards correlations between particles. Using a different mathematical approach might lead to possible analytical results for these correlations.\\

\begin{acknowledgments}
The authors would like to thank Tom Williams for a critical reading of this paper.  We would also like to thank the anonymous referees for their valuable feedback.  Funding for this research was provided through the Summer Research Scholars Program and the Lenfest Grant at Washington and Lee University.
\end{acknowledgments}

\end{document}